\documentclass[12pt]{iopart}
\usepackage{iopams}  
\usepackage{graphicx}
\usepackage{dcolumn}
\usepackage{amsfonts}
\usepackage{latexsym}
\usepackage{amssymb} 
\usepackage{verbatim}
\usepackage{amsthm}

\newcommand{\RF}{{{\mathbb R}}}

\newtheorem{conjecture}{Conjecture}
\begin{document}
\title[
  General formulation of higher-order
  gauge-invariant perturbation theory 
]{
  General formulation of general-relativistic higher-order 
  gauge-invariant perturbation theory
}
\author{
  Kouji Nakamura
}
\address{
  Optical and Infrared Astronomy Division,\\
  National Astronomical Observatory of Japan,\\
  2-21-1, Osawa, Mitaka, Tokyo 181-8588, Japan
}
\ead{kouji.nakamura@nao.ac.jp}
\begin{abstract}
  Gauge-invariant treatments of general-relativistic
  higher-order perturbations on generic background spacetime is
  proposed.
  After reviewing the general framework of the second-order
  gauge-invariant perturbation theory, we show the fact that the
  linear-order metric perturbation is decomposed into
  gauge-invariant and gauge-variant parts, which was the
  important premis of this general framework.
  This means that the development the higher-order
  gauge-invariant perturbation theory on generic background
  spacetime is possible.
  A remaining issue to be resolve is also disscussed.
\end{abstract}


\section{Introduction}
\label{sec:introduction}


Perturbation theories are powerful techniques in many area of
physics and lead phyisically fruiteful results.
In particualr, in general relativity, the construction of exact
soltuions is not so easy and known exact solutions are often too
idealized, though there are many known exact solutions to the
Einstein
equation\cite{H.Stephani-D.Kramer-M.A.N.MacCallum-C.Hoenselaers-E.Herlt-2003}. 
Of course, some exact solutions to the Einstein equation well
describe our universe, gravitational field of stars, and black
holes.
However, in natural phenomena, there always exist
``fluctuations''. 
To describe these fluctuations, the {\it linear} perturbation
theories around some background spacetime are
developed\cite{Bardeen-1980}, 
and are used to describe fluctuations of our universe,
gravitational field of stars, and gravitational waves
from strongly gravitating sources.


Besides the development of the general-relativistic linear-order
perturbation theory, {\it higher-order} general-relativistic
perturbations also have very wide applications, for example,
cosmological 
perturbations\cite{Tomita-1967,M.Bruni-S.Soonego-CQG1997,S.Sonego-M.Bruni-CMP1998,kouchan-cosmo-second,kouchan-second-cosmo-matter},
black hole perturbations\cite{Gleiser-Nicasio}, and perturbation
of a neutron star\cite{Kojima}.
In spite of these wide applications, there is a delicate issue
in the treatment of general-relativistic perturbations, which is
called {\it gauge issue}.
General relativity is based on general covariance.
Due to this general covariance, the {\it gauge degree of
  freedom}, which is an unphysical degree of freedom of
perturbations, arises in general-relativistic perturbations.
To obtain physical results, we have to fix this gauge degree
of freedom or to treat some invariant quantities in
perturbations. 
This situation becomes more complicated in higher-order
perturbations.
Therefore, it is worthwhile to investigate higher-order
gauge-invariant perturbation theory from a general point of
view.


According to this motivation, the general framework of
higher-order general-relativistic gauge-invariant perturbation 
theory has been discussed\cite{kouchan-gauge-inv,kouchan-second}
and applied to cosmological
perturbations\cite{kouchan-cosmo-second,kouchan-second-cosmo-matter}.
However, this framework is based on a conjecture (Conjecture
\ref{conjecture:decomposition-conjecture} below) which roughly
states that {\it we have already known the procedure to find 
  gauge-invariant variables for a linear-order metric
  perturbations}.
The main purpose of this letter is to give the outline of a
proof of this conjecture.
Due to this proof, a formulation of the higher-order
general-relativistic gauge-invariant perturbation theory is
almost completed on generic background spacetime.
The details of the ingredients of this Letter are explained in
Ref.~\cite{kouchan-in-preparation}.


\section{General framework of higher-order gauge-invariant perturbation theory}
\label{sec:General-framework-of-the-gauge-invariant-perturbation-theory}


In this section, we review the framework of the gauge-invariant
perturbation theory\cite{kouchan-gauge-inv,kouchan-second}.
In any perturbation theory, we always treat two spacetime
manifolds.
One is the physical spacetime $({\cal M},\bar{g}_{ab})$, which
is our nature itself, and we want to describe 
$({\cal M},\bar{g}_{ab})$ by perturbations. 
The other is the background spacetime $({\cal M}_{0},g_{ab})$,
which is prepared as a reference by hand.
We note that these two spacetimes are distinct.


Further, in any perturbation theory, we always write equations 
for the perturbation of the variable $Q$ like
\begin{equation}
  \label{eq:variable-symbolic-perturbation}
  Q(``p\mbox{''}) = Q_{0}(p) + \delta Q(p).
\end{equation}
Equation (\ref{eq:variable-symbolic-perturbation}) gives a
relation between variables on different manifolds.
Actually, $Q(``p\mbox{''})$ in
Eq.~(\ref{eq:variable-symbolic-perturbation}) is a variable on 
${\cal M}$, while $Q_{0}(p)$ and $\delta Q(p)$ are variables on
${\cal M}_{0}$.
Since we regard Eq.~(\ref{eq:variable-symbolic-perturbation}) as
a field equation, Eq.~(\ref{eq:variable-symbolic-perturbation})
includes an implicit assumption of the exitence of a point
identification map ${\cal M}_{0}\rightarrow{\cal M}$ $:$
$p\in{\cal M}_{0}\mapsto ``p\mbox{''}\in{\cal M}$.
This idenification map is a {\it gauge choice} in
perturbation theories\cite{J.M.Stewart-M.Walker11974}.


To develop this understanding of the ``gauge'', we introduce an
infinitesimal parameter $\lambda$ for perturbations and
$(n+1)+1$-dimensional manifold ${\cal N}={\cal M}\times\RF$
($n+1=\dim{\cal M}$) so that 
${\cal M}_{0}=\left.{\cal N}\right|_{\lambda=0}$ and
${\cal M}={\cal M}_{\lambda}=\left.{\cal N}\right|_{\RF=\lambda}$.
On ${\cal N}$, the gauge choice is regarded as a diffeomorphism  
${\cal X}_{\lambda}:{\cal N}\rightarrow{\cal N}$ such that
${\cal X}_{\lambda}:{\cal M}_{0}\rightarrow{\cal M}_{\lambda}$.
Further, we introduce a gauge choice ${\cal X}_{\lambda}$ as an
exponential map with a generator ${}^{{\cal X}}\!\eta^{a}$
which is chosen so that its integral curve in ${\cal N}$ is
transverse to each ${\cal M}_{\lambda}$ everywhere on 
${\cal N}$\cite{M.Bruni-S.Soonego-CQG1997,S.Sonego-M.Bruni-CMP1998}. 
Points lying on the same integral curve are regarded as the
``same'' by the gauge choice ${\cal X}_{\lambda}$.


The first- and the second-order perturbations of
the variable $Q$ on ${\cal M}_{\lambda}$ are defined by the
pulled-back ${\cal X}_{\lambda}^{*}Q$ on ${\cal M}_{0}$, which
is induced by ${\cal X}_{\lambda}$, and expanded as
\begin{eqnarray}
  {\cal X}_{\lambda}^{*}Q
  =
  Q_{0}
  + \lambda \left.{\pounds}_{{}^{{\cal X}}\!\eta}Q\right|_{{\cal M}_{0}}
  + \frac{1}{2} \lambda^{2} 
  \left.{\pounds}_{{}^{{\cal X}}\!\eta}^{2}Q\right|_{{\cal M}_{0}}
  + O(\lambda^{3}),
  \label{eq:perturbative-expansion-of-Q-def}
\end{eqnarray}
$Q_{0}=\left.Q\right|_{{\cal M}_{0}}$ is the background value of
$Q$ and all terms in
Eq.~(\ref{eq:perturbative-expansion-of-Q-def}) are evaluated on
${\cal M}_{0}$.
Since Eq.~(\ref{eq:perturbative-expansion-of-Q-def}) is just the
perturbative expansion of ${\cal X}^{*}_{\lambda}Q_{\lambda}$,
the first- and the second-order perturbations of $Q$ are given
by  
${}^{(1)}_{{\cal X}}\!Q:=\left.{\pounds}_{{}^{{\cal X}}\!\eta}Q\right|_{{\cal M}_{0}}$
and 
${}^{(2)}_{{\cal X}}\!Q:=\left.{\pounds}_{{}^{{\cal X}}\!\eta}^{2}Q\right|_{{\cal M}_{0}}$,
respectively.


When we have two gauge choices ${\cal X}_{\lambda}$ and 
${\cal Y}_{\lambda}$ with the generators ${}^{{\cal X}}\!\eta^{a}$
and ${}^{{\cal Y}}\!\eta^{a}$, respectively, and when these
generators have the different tangential componetns 
to each ${\cal M}_{\lambda}$, ${\cal X}_{\lambda}$ and 
${\cal Y}_{\lambda}$ are regarded as {\it different gauge choices}.
The {\it gauge-transformation} is regarded as the change of the
gauge choice ${\cal X}_{\lambda}\rightarrow{\cal Y}_{\lambda}$,
which is given by the diffeomorphism 
$\Phi_{\lambda}:=\left({\cal X}_{\lambda}\right)^{-1}\circ{\cal Y}_{\lambda}
  : {\cal M}_{0} \rightarrow {\cal M}_{0}$.
The diffeomorphism $\Phi_{\lambda}$ does change the point
identification.
$\Phi_{\lambda}$ induces a pull-back from the representation
${\cal X}_{\lambda}^{*}\!Q_{\lambda}$ to the representation
${\cal Y}_{\lambda}^{*}\!Q_{\lambda}$ as 
${\cal Y}_{\lambda}^{*}\!Q_{\lambda}=\Phi_{\lambda}^{*}{\cal X}_{\lambda}^{*}\!Q_{\lambda}$.
From general arguments of the Taylor
expansion\cite{S.Sonego-M.Bruni-CMP1998}, the pull-back
$\Phi_{\lambda}^{*}$ is expanded as
\begin{eqnarray}
  {\cal Y}_{\lambda}^{*}\!Q_{\lambda}
  &=&
  {\cal X}_{\lambda}^{*}\!Q_{\lambda}
  + \lambda {\pounds}_{\xi_{(1)}} {\cal X}_{\lambda}^{*}\!Q_{\lambda}
  + \frac{1}{2} \lambda \left(
    {\pounds}_{\xi_{(2)}} + {\pounds}_{\xi_{(1)}}^{2}
  \right) {\cal X}_{\lambda}^{*}\!Q_{\lambda}
  + O(\lambda^{3}),
  \label{eq:Bruni-46-one}
\end{eqnarray}
where $\xi_{(1)}^{a}$ and $\xi_{(2)}^{a}$ are the genertors of
$\Phi_{\lambda}$.
From Eqs.~(\ref{eq:perturbative-expansion-of-Q-def}) and
(\ref{eq:Bruni-46-one}), each order gauge-transformation is
given as
\begin{eqnarray}
  \label{eq:Bruni-47-one}
  {}^{(1)}_{\;{\cal Y}}\!Q - {}^{(1)}_{\;{\cal X}}\!Q &=& 
  {\pounds}_{\xi_{(1)}}Q_{0}, \\
  \label{eq:Bruni-49-one}
  {}^{(2)}_{\;\cal Y}\!Q - {}^{(2)}_{\;\cal X}\!Q &=& 
  2 {\pounds}_{\xi_{(1)}} {}^{(1)}_{\;\cal X}\!Q 
  +\left\{{\pounds}_{\xi_{(2)}}+{\pounds}_{\xi_{(1)}}^{2}\right\} Q_{0}.
\end{eqnarray}
We also employ the {\it order by order gauge invariance} as a
concept of gauge invariance\cite{kouchan-second-cosmo-matter}. 
We call the $k$th-order perturbation ${}^{(p)}_{{\cal X}}\!Q$ is
gauge invariant iff ${}^{(k)}_{\;\cal X}\!Q = {}^{(k)}_{\;\cal Y}\!Q$
for any gauge choice ${\cal X}_{\lambda}$ and
${\cal Y}_{\lambda}$.


Based on the above set up, we proposed a procedure to construct
gauge-invariant variables of higher-order
perturbations\cite{kouchan-gauge-inv}.
First, we expand the metric on the physical spacetime 
${\cal M}_{\lambda}$, which is pulled back to the background
spacetime ${\cal M}_{0}$ through a gauge choice 
${\cal X}_{\lambda}$ as 
\begin{eqnarray}
  {\cal X}^{*}_{\lambda}\bar{g}_{ab}
  &=&
  g_{ab} + \lambda {}_{{\cal X}}\!h_{ab} 
  + \frac{\lambda^{2}}{2} {}_{{\cal X}}\!l_{ab}
  + O^{3}(\lambda).
  \label{eq:metric-expansion}
\end{eqnarray}
Although the expression (\ref{eq:metric-expansion}) depends
entirely on the gauge choice ${\cal X}_{\lambda}$, henceforth,
we do not explicitly express the index of the gauge choice
${\cal X}_{\lambda}$ in the expression if there is no
possibility of confusion. 
The important premise of our proposal was the following
conjecture\cite{kouchan-gauge-inv} for $h_{ab}$ : 
\begin{conjecture}
  \label{conjecture:decomposition-conjecture}
  For a second-rank tensor $h_{ab}$, whose gauge transformation
  is given by (\ref{eq:Bruni-47-one}), there exist a tensor
  ${\cal H}_{ab}$ and a vector $X^{a}$ such that $h_{ab}$ is
  decomposed as
  \begin{eqnarray}
    h_{ab} =: {\cal H}_{ab} + {\pounds}_{X}g_{ab},
    \label{eq:linear-metric-decomp}
  \end{eqnarray}
  where ${\cal H}_{ab}$ and $X^{a}$ are transformed as
  \begin{equation}
    {}_{{\cal Y}}\!{\cal H}_{ab} - {}_{{\cal X}}\!{\cal H}_{ab} =  0, 
    \quad
    {}_{\quad{\cal Y}}\!X^{a} - {}_{{\cal X}}\!X^{a} = \xi^{a}_{(1)} 
    \label{eq:linear-metric-decomp-gauge-trans}
  \end{equation}
  under the gauge transformation (\ref{eq:Bruni-47-one}),
  respectively.
\end{conjecture}
We call ${\cal H}_{ab}$ and $X^{a}$ are the 
{\it gauge-invariant part} and the {\it gauge-variant part} 
of $h_{ab}$, respectively.


Although Conjecture \ref{conjecture:decomposition-conjecture} is
nontrivial on generic background spacetime, once we accept this
conjecture, we can always find gauge-invariant variables for
higher-order perturbations\cite{kouchan-gauge-inv}. 
Using Conjecture \ref{conjecture:decomposition-conjecture}, the
second-order metric perturbation $l_{ab}$ is decomposed as
\begin{eqnarray}
  \label{eq:H-ab-in-gauge-X-def-second-1}
  l_{ab}
  =:
  {\cal L}_{ab} + 2 {\pounds}_{X} h_{ab}
  + \left(
      {\pounds}_{Y}
    - {\pounds}_{X}^{2} 
  \right)
  g_{ab},
\end{eqnarray}
where ${}_{{\cal Y}}\!{\cal L}_{ab}-{}_{{\cal X}}\!{\cal L}_{ab}=0$  
and 
${}_{{\cal Y}}\!Y^{a}-{}_{{\cal X}}\!Y^{a}=\xi_{(2)}^{a}+[\xi_{(1)},X]^{a}$.
Furthermore, using the first- and second-order gauge-variant
parts, $X^{a}$ and $Y^{a}$, of the metric perturbations,
gauge-invariant variables for an arbitrary tensor field $Q$ 
other than the metric can be defined by
\begin{eqnarray}
  \label{eq:matter-gauge-inv-def-1.0}
  {}^{(1)}\!{\cal Q} &:=& {}^{(1)}\!Q - {\pounds}_{X}Q_{0}
  , \\ 
  \label{eq:matter-gauge-inv-def-2.0}
  {}^{(2)}\!{\cal Q} &:=& {}^{(2)}\!Q - 2 {\pounds}_{X} {}^{(1)}Q 
  - \left\{ {\pounds}_{Y} - {\pounds}_{X}^{2} \right\} Q_{0}
  .
\end{eqnarray}
These definitions (\ref{eq:matter-gauge-inv-def-1.0}) and
(\ref{eq:matter-gauge-inv-def-2.0}) also imply that any
perturbation of first and second order is always decomposed
into gauge-invariant and gauge-variant parts as
\begin{eqnarray}
  \label{eq:matter-gauge-inv-decomp-1.0}
  {}^{(1)}\!Q &=& {}^{(1)}\!{\cal Q} + {\pounds}_{X}Q_{0}
  , \\ 
  \label{eq:matter-gauge-inv-decomp-2.0}
  {}^{(2)}\!Q  &=& {}^{(2)}\!{\cal Q} + 2 {\pounds}_{X} {}^{(1)}Q 
  + \left\{ {\pounds}_{Y} - {\pounds}_{X}^{2} \right\} Q_{0}
  ,
\end{eqnarray}
respectively.


Actually, the perturbations of the Einstein tensor are given in
the same form Eqs.~(\ref{eq:matter-gauge-inv-decomp-1.0}) and
(\ref{eq:matter-gauge-inv-decomp-2.0}) :
\begin{eqnarray}
  \bar{G}_{a}^{\;\;b}
  &=&
  G_{a}^{\;\;b}
  + \lambda {}^{(1)}\!G_{a}^{\;\;b} 
  + \frac{1}{2} \lambda^{2} {}^{(2)}\!G_{a}^{\;\;b} 
  + O(\lambda^{3})
  , \\
  \label{eq:linear-Einstein}
  {}^{(1)}\!G_{a}^{\;\;b}
  &=&
  {}^{(1)}{\cal G}_{a}^{\;\;b}\left[{\cal H}\right]
  + {\pounds}_{X} G_{a}^{\;\;b}
  ,\\
  {}^{(2)}\!G_{a}^{\;\;b}
  &=& 
  {}^{(1)}{\cal G}_{a}^{\;\;b}\left[{\cal L}\right]
  + {}^{(2)}{\cal G}_{a}^{\;\;b} \left[{\cal H}, {\cal H}\right]
  + 2 {\pounds}_{X} {}^{(1)}\!\bar{G}_{a}^{\;\;b}
  + \left\{ {\pounds}_{Y} - {\pounds}_{X}^{2} \right\} G_{a}^{\;\;b},
  \label{eq:second-Einstein-2,0-0,2}
\end{eqnarray}
where ${}^{(1)}{\cal G}_{a}^{\;\;b}\left[*\right]$ is the
gauge-invariant linear terms and 
${}^{(2)}{\cal G}_{a}^{\;\;b}\left[*,*\right]$ are collections
of quadratic terms of gauge-invaraint linear metric
perturbations.
On the other hand, the energy momentum tensor on 
${\cal M}_{\lambda}$ is also expanded as 
\begin{eqnarray}
  \bar{T}_{a}^{\;\;b}
  =
  T_{a}^{\;\;b}
  + \lambda {}^{(1)}\!T_{a}^{\;\;b}
  + \frac{1}{2} \lambda^{2} {}^{(2)}\!T_{a}^{\;\;b}
  + O(\lambda^{3}),
\end{eqnarray}
and its first- and the second-order perturbations
${}^{(1)}\!T_{a}^{\;\;b}$ and ${}^{(2)}\!T_{a}^{\;\;b}$ are
decomposed as Eqs.~(\ref{eq:matter-gauge-inv-decomp-1.0}) and 
(\ref{eq:matter-gauge-inv-decomp-2.0}) :
\begin{eqnarray}
  {}^{(1)}\!T_{a}^{\;\;b}
  &=&
  {}^{(1)}\!{\cal T}_{a}^{\;\;b}
  + {\pounds}_{X}T_{a}^{\;\;b}
  , \\
  {}^{(2)}\!T_{a}^{\;\;b}
  &=&
  {}^{(2)}\!{\cal T}_{a}^{\;\;b}
  + 2 {\pounds}_{X}{}^{(1)}\!T_{a}^{\;\;b}
  + \left\{ {\pounds}_{Y} - {\pounds}_{X}^{2} \right\} T_{a}^{\;\;b}
  .
\end{eqnarray}
These were confirmed in the case of a perfect fluid, an
imperfect fluid, and a scalar
field\cite{kouchan-second-cosmo-matter}.


Imposing order by order Einstein equations
\begin{eqnarray}
  G_{a}^{\;\;b} = 8\pi T_{a}^{\;\;b}, \quad
  {}^{(p)}\!G_{a}^{\;\;b} = 8\pi {}^{(p)}\!T_{a}^{\;\;b}, \quad
  (p=1,2),
\end{eqnarray}
the first- and the second-order perturbation of the Einstein
equations are automatically given in gauge-invariant form :
\begin{eqnarray}
  {}^{(1)}\!{\cal G}_{a}^{\;\;b}\left[{\cal H}\right]
  =
  8\pi G {}^{(1)}{\cal T}_{a}^{\;\;b}
  ,
  \quad
  {}^{(1)}\!{\cal G}_{a}^{\;\;b}\left[{\cal L}\right]
  + {}^{(2)}\!{\cal G}_{a}^{\;\;b}\left[{\cal H}, {\cal H}\right]
  =
  8\pi G \;\; {}^{(2)}{\cal T}_{a}^{\;\;b} .
\end{eqnarray}
Further, the perturbative equations of motion for matter
fields, which are derived from the divergence of the energy
momentum tensor, are also automatically given in gauge-invariant 
form\cite{kouchan-second-cosmo-matter}.


Thus, based only on Conjecture
\ref{conjecture:decomposition-conjecture}, we have developed
the general framework of second-order general relativistic
perturbation theory. 
We also note that this general framework of the second-order
gauge-invariant perturbation theory are independent of the
explicit form of the background metric $g_{ab}$, except 
for Conjecture \ref{conjecture:decomposition-conjecture}.


\section{Decomposition of the linear-order metric perturbation}
\label{sec:K.Nakamura-2010-3}


Now, we give the outline of a proof of Conjecture
\ref{conjecture:decomposition-conjecture}.
To do this, we only consider the background spacetimes which
admit ADM decomposition\cite{wald-book}.
Therefore, the background spacetime ${\cal M}_{0}$ considered
here is $n+1$-dimensional spacetime which is desribed by the
direct product $\RF\times\Sigma$.
Here, $\RF$ is a time direction and $\Sigma$ is the spacelike
hypersurface ($\dim\Sigma = n$).
The background metric $g_{ab}$ is given as
\begin{eqnarray}
  \label{eq:gdb-decomp-dd-minus-main}
  g_{ab} \! = \! - \alpha^{2} (dt)_{a} (dt)_{b}
  + q_{ij}
  (dx^{i} + \beta^{i}dt)_{a}
  (dx^{j} + \beta^{j}dt)_{b}.
\end{eqnarray}
In this letter, we only consider the case where
$\alpha = 1$ and $\beta^{i} = 0$, for simlicity.
The most general case where $\alpha\neq 1$ and
$\beta^{i}\neq 0$ is discussed in
Ref.~\cite{kouchan-in-preparation}.


To consider the decomposition (\ref{eq:linear-metric-decomp}) of
$h_{ab}$, first, we consider the components of the metric
$h_{ab}$ as 
\begin{eqnarray}
  \label{eq:hab-ADM-decomp}
  h_{ab}
  &=&
  h_{tt} (dt)_{a}(dt)_{b}
  + 2 h_{ti} (dt)_{(a}(dx^{i})_{b)}
  + h_{ij} (dx^{i})_{a}(dx^{j})_{b}.
\end{eqnarray}
Under the gauge-transformation (\ref{eq:Bruni-47-one}), in the
case where $\alpha=1$ and $\beta^{i}=0$, these components
$\{h_{tt},h_{ti},h_{ij}\}$ are transformed as 
\begin{eqnarray}
  \label{eq:gauge-trans-of-htt-ADM-BG-case2}
  {}_{{\cal Y}}h_{tt}
  -
  {}_{{\cal X}}h_{tt}
  &=&
  2 \partial_{t}\xi_{t}
  , \\
  \label{eq:gauge-trans-of-hti-ADM-BG-case2}
  {}_{{\cal Y}}h_{ti}
  -
  {}_{{\cal X}}h_{ti}
  &=&
  \partial_{t}\xi_{i}
  + D_{i}\xi_{t}
  + 2 K^{j}_{\;\;i} \xi_{j}
  , \\
  \label{eq:gauge-trans-of-hij-ADM-BG-case2}
  {}_{{\cal Y}}h_{ij}
  -
  {}_{{\cal X}}h_{ij}
  &=&
  2 D_{(i}\xi_{j)}
  + 2 K_{ij} \xi_{t}
  .
\end{eqnarray}
where $K_{ij}$ is the extrinsic curvature of $\Sigma$ and
$D_{i}$ is the covariant derivative associate with the metric
$q_{ij}$ ($D_{i}q_{jk}=0$). 
In our case, $K_{ij}=-\frac{1}{2} \partial_{t}q_{ij}$.


Inspecting gauge-transformation rules
(\ref{eq:gauge-trans-of-hti-ADM-BG-case2}) and
(\ref{eq:gauge-trans-of-hij-ADM-BG-case2}), 
we introduce a new symmetric tensor $\hat{H}_{ab}$ whose
components are given by 
\begin{eqnarray}
  \hat{H}_{tt} := h_{tt}, \quad
  \hat{H}_{ti} := h_{ti}, \quad
  \hat{H}_{ij} := h_{ij} - 2 K_{ij}X_{t}.
  \label{eq:hatH-def-case2}
\end{eqnarray}
Here, we assume the existence of the variable $X_{t}$
whose gauge-transformation rule is given by 
${}_{{\cal Y}}X_{t}-{}_{{\cal X}}X_{t}=\xi_{t}$.
This assumption is confirmed later soon. 
Since the components $\hat{H}_{ti}$ and $\hat{H}_{ij}$ are
regarded as a vector and a symmetric tensor on $\Sigma$,
respectively, $\hat{H}_{ti}$ and $\hat{H}_{ij}$ are decomposed
as\cite{J.W.York-1973}
\begin{eqnarray}
  \label{eq:K.Nakamura-2010-2-simple-4-7}
  \hat{H}_{ti} &=& D_{i}h_{(VL)} + h_{(V)i}, \quad D^{i}h_{(V)i} = 0,
  \\
  \label{eq:K.Nakamura-2010-2-simple-4-8}
  \hat{H}_{ij} &=& \frac{1}{n} q_{ij} h_{(L)}
  + 2 \left(D_{(i}h_{(TV)j)} - \frac{1}{n}q_{ij}D^{l}h_{(TV)l}\right)
  + h_{(TT)ij},
  \\
  \label{eq:K.Nakamura-2010-2-simple-4-10}
  h_{(TV)i} &=& D_{i}h_{(TVL)} + h_{(TVV)i}, \quad
  D^{i}h_{(TVV)i} = 0, \quad
  D^{i}h_{(TT)ij} = 0.
\end{eqnarray}
To confirm the one-to-one correspondence between
$\{\hat{H}_{ti}$, $\hat{H}_{ij}\}$ and $\{h_{(VL)}$, $h_{(V)i}$,
$h_{(L)}$, $h_{(TVL)}$, $h_{(TVV)i}$, $h_{(TT)ij}\}$, we have to
discuss the boundary conditions for the variables $\{h_{(VL)}$, 
$h_{(V)i}$, $h_{(L)}$, $h_{(TVL)}$, $h_{(TVV)i}$, $h_{(TT)ij}\}$
on the background hypersurface $(\Sigma,q_{ab})$.
However, in this paper, we do not discuss these boundary
conditions in detail.
Instead, we assume the existence of the Green functions of
two elliptic derivative operators $\Delta:=D^{i}D_{i}$ and  
${\cal D}^{ij}:=q^{ij}\Delta+\left(1-\frac{2}{n}\right)D^{i}D^{j}+{}^{(n)}\!R^{ij}$,
where ${}^{(n)}\!R^{ij}$ is the Ricci curvature on $\Sigma$.
The boundary conditions for the variables are implicitly
included in the Green functions of these derivative operator and
the one-to-one correspondence of the sets $\{\hat{H}_{ti}$,
$\hat{H}_{ij}\}$ and $\{h_{(VL)}$, $h_{(V)i}$, $h_{(L)}$,
$h_{(TVL)}$, $h_{(TVV)i}$, $h_{(TT)ij}\}$ are guaranteed by
these two Green functions.
Further, we also decompose the component $\xi_{i}$ of the
generator of gauge transformation as
$\xi_{i}=:D_{i}\xi_{(L)}+\xi_{(V)i}$. 
Gauge-transformation rules for $\{h_{tt}$, $h_{(VL)}$,
$h_{(V)i}$, $h_{(L)}$, $h_{(TVL)}$, $h_{(TVV)i}$, $h_{(TT)ij}\}$ 
are summarized as
\begin{eqnarray}
  {}_{{\cal Y}}h_{tt}
  -
  {}_{{\cal X}}h_{tt}
  &=&
  2 \partial_{t}\xi_{t}
  \label{eq:K.Nakamura-2010-2-simple-4-39-2}
  , \\
  {}_{{\cal Y}}h_{(VL)} - {}_{{\cal X}}h_{(VL)}
  &=&
    \partial_{t}\xi_{(L)}
  + \xi_{t}
  \nonumber\\
  && \quad
  +
  \Delta^{-1}
  \left[
    2 D_{i}\left( K^{ij} D_{j}\xi_{(L)} \right)
    + D^{k}K \xi_{(V)k}
  \right]
  \label{eq:K.Nakamura-2010-2-simple-4-40}
  , \\
  {}_{{\cal Y}}h_{(V)i}
  -
  {}_{{\cal X}}h_{(V)i}
  &=&
    \partial_{t}\xi_{(V)i}
  + 2 K^{j}_{\;\;i} D_{j}\xi_{(L)}
  + 2 K^{j}_{\;\;i} \xi_{(V)j}
  \nonumber\\
  && \quad
  - D_{i}\Delta^{-1}
  \left[
    2 D_{k} \left( K^{kj} D_{j}\xi_{(L)} \right)
    + D^{k}K \xi_{(V)k}
  \right]
  \label{eq:K.Nakamura-2010-2-simple-4-41}
  ,
  \\
  {}_{{\cal Y}}h_{(L)}
  -
  {}_{{\cal X}}h_{(L)}
  &=&
  2 D^{i}\xi_{i}
  \label{eq:K.Nakamura-2010-2-simple-4-42}
  , \quad
  \\
  \label{eq:K.Nakamura-2010-2-simple-4-45}
  {}_{{\cal Y}}h_{(TVL)} - {}_{{\cal X}}h_{(TVL)}
  &=&
  \xi_{(L)}
  ,
  \\
  \label{eq:K.Nakamura-2010-2-simple-4-46}
  {}_{{\cal Y}}h_{(TVV)l} - {}_{{\cal X}}h_{(TVV)l}
  &=&
  \xi_{(V)l}
  , \\
  {}_{{\cal Y}}h_{(TT)ij} - {}_{{\cal X}}h_{(TT)ij}
  &=&
  0
  .
  \label{eq:K.Nakamura-2010-2-simple-4-44}
\end{eqnarray}


We first find the variable $X_{t}$ in Eq.~(\ref{eq:hatH-def-case2}).
From the above gauge-transformation rules, we see that
the combination 
\begin{eqnarray}
  X_{t}
  &:=&
  h_{(VL)}
  - \partial_{t}h_{(TVL)}
  -
  \Delta^{-1}
  \left[
      2 D_{k}\left( K^{kj}D_{j}h_{(TVL)} \right)
    +   D^{k}K h_{(TVV)k}
  \right]
  \label{eq:K.Nakamura-2010-2-simple-4-53}
\end{eqnarray}
satisfy ${}_{{\cal Y}}X_{t}-{}_{{\cal X}}X_{t}=\xi_{t}$. 
We also find the variable $X_{i}$
\begin{eqnarray}
  \label{eq:K.Nakamura-2010-2-simple-4-56}
  X_{i} := h_{(TV)i} = D_{i}h_{(TVL)} + h_{(TVV)i}
\end{eqnarray}
satisfy the gauge-transformation rule 
${}_{{\cal Y}}X_{i}-{}_{{\cal X}}X_{i}=\xi_{i}$.


Inspecting gauge-transformation rules
(\ref{eq:K.Nakamura-2010-2-simple-4-39-2})--(\ref{eq:K.Nakamura-2010-2-simple-4-44})
and using the variables $X_{t}$ and $X_{i}$ defined by
Eqs.~(\ref{eq:K.Nakamura-2010-2-simple-4-53})--(\ref{eq:K.Nakamura-2010-2-simple-4-56}), 
we find gauge-invariant variables as follows:
\begin{eqnarray}
  \label{eq:K.Nakamura-2010-2-simple-4-58}
  - 2 \Phi &:=& h_{tt} - 2 \partial_{t}\hat{X}_{t}, \\
  - 2 n \Psi &:=& h_{(L)} - 2 D^{i}\hat{X}_{i}, \\
  \nu_{i}
  &:=&
  h_{(V)i}
  - \partial_{t}h_{(TVV)i}
  - 2 K^{j}_{\;\;i} \left( D_{j}h_{(TVL)} + h_{(TVV)j} \right)
  \nonumber\\
  &&
  + D_{i}\Delta^{-1}
  \left[
    2 D_{k}\left( K^{kj} D_{j}h_{(TVL)} \right)
    + D^{k}K h_{(TVV)k}
  \right]
  \label{eq:K.Nakamura-2010-2-simple-4-48}
  , \\
  \chi_{ij} &:=& h_{(TT)ij}.
\end{eqnarray}
Actually, it is straightforward to confirm the gauge-invariance
of these varaibles.


In terms of the variables $\Phi$, $\Psi$, $\nu_{i}$,
$\chi_{ij}$, $X_{t}$, and $X_{i}$, original components of
$h_{ab}$ is given by 
\begin{eqnarray}
  \label{eq:K.Nakamura-2010-2-simple-4-71}
  h_{tt} &=& - 2 \Phi + 2 \partial_{t}X_{t}, \\
  h_{ti}
  &=&
    \nu_{i}
  + D_{i}X_{t}
  + \partial_{t}X_{i}
  + 2 K^{j}_{\;\;i} X_{j}
  \label{eq:K.Nakamura-2010-2-simple-4-72}
  , \\
  h_{ij}
  &=&
  - 2 \Psi q_{ij} 
  + \chi_{ij}
  + D_{i}X_{j} + D_{j}X_{i}
  + 2 K_{ij} X_{t}
  \label{eq:K.Nakamura-2010-2-simple-4-73}
  .
\end{eqnarray}
Comparing Eq.~(\ref{eq:linear-metric-decomp}),
a natural choice of ${\cal H}_{ab}$ and $X_{a}$ are
\begin{eqnarray}
  \label{eq:calHab-component-identification-case2}
  {\cal H}_{ab}
  &=& - 2 \Phi (dt)_{a}(dt)_{b}
  + 2 \nu_{i} (dt)_{(a}(dx^{i})_{b)}
  + \left(- 2 \Psi q_{ij} + \chi_{ij}\right) (dx^{i})_{a} (dx^{i})_{b}
  , \\ 
  \label{eq:Xt-Xi-component-identification-case2}
  X_{a} &=& X_{t}(dt)_{a} + X_{i} (dx^{i}).
\end{eqnarray}
These show that the linear-order metric perturbation
$h_{ab}$ is decomposed into the form
Eq.~(\ref{eq:linear-metric-decomp}).


\section{Summary and discussions}
\label{sec:summary}


In summary, we showed the outline of a proof of Conjecture
\ref{conjecture:decomposition-conjecture} which is the important 
premise of our general framework of gauge-invariant perturbation
theory.
Although we only consider the background spacetime with
$\alpha=1$ and $\beta^{i}=0$, the above proof is extended to
general case where $\alpha\neq 1$ and 
$\beta^{i}\neq 0$\cite{kouchan-in-preparation}.
Further, in our proposal, we applied the ADM decomposition to
identify the gauge-ivariant variable ${\cal H}_{ab}$ and the
gauge-variant variable $X_{a}$, and our choice of the components 
of ${\cal H}_{ab}$ depends on the choice of the spacelike
hypersurface $\Sigma$ in ${\cal M}_{0}$. 
However, this $\Sigma$ dependence of our choice does not break
the covariance in the statement of Conjecture
\ref{conjecture:decomposition-conjecture}.
This can be easily see from the extension of our proof to
the most general case where $\alpha\neq 1$ and $\beta^{i}\neq 0$
in Ref.~\cite{kouchan-in-preparation}.
We also note that the choice the decomosition
(\ref{eq:linear-metric-decomp}) is not unique as pointed out in
Ref.~\cite{kouchan-second-cosmo-matter}.
What we showed is a procedure to carry out the decomposition
(\ref{eq:linear-metric-decomp}) to emphasize the existence of 
${\cal H}_{ab}$ and $X_{a}$ in
Eq.~(\ref{eq:linear-metric-decomp}).


In our proof, we assumed the existence of the Green functions for
the derivative operators $\Delta$ and ${\cal D}^{ij}$.
In this sense, we have specified the boundary conditions for the 
perturbative variables at the boundary $\partial\Sigma$ of
$\Sigma$, because explicit expression of Green functions are
depends on the boundary conditions.
This also implies that we have ingored the ``zero-mode'' which
belong to the kernel of these derivative operators.
These zero-mode corresponds to the degree of freefom of the
boundary condition at the boundary $\partial\Sigma$.
To includes these modes into our consideration, different
treatments perturbations and the careful arguments of the
boundary conditions for the perturbations will be necessary.
We call this problem as {\it zero-mode problem}.
Even in the cosmological perturbations, zero-mode problem
exists.
We leave the resolution of this zero-mode problem as a future
work.


Although this zero-mode problem should be resolved, we confirmed 
the important premise of our general framework of second-order
gauge-invaraint perturbation theory on generic background
spacetime.
Due to this, we have the possibility of applications of our
framework for the second-order gauge-invariant perturbation
theory to perturbations on generic background spacetime. 
Actually, in the cosmological perturbation case, we have
developed the second-order cosmological perturbations along 
this general framework\cite{kouchan-cosmo-second,kouchan-second-cosmo-matter}.
The similar development will be also possible for the any order
perturbation in two-parameter case\cite{kouchan-gauge-inv}.
Therefore, we may say that the wide appicaltions of our
gauge-invariant perturbation theory are opened.
We also leave these developments as future works.


\section*{Acknowledgements}


The author deeply acknowledged to Professor Masa-Katsu Fujimoto
in National Astronomical Observatory of Japan for his various
support.


\section*{References}

\end{document}